\documentclass[aps,prl,twocolumn,groupedaddress,showpacs,letterpaper,floatfix,10pt]{revtex4-1}
\usepackage{amsmath,amssymb,graphicx,hyperref}
\hypersetup{pdfnewwindow=true, colorlinks=true, linkcolor=blue, anchorcolor=blue,
  citecolor=blue, filecolor=blue, menucolor=blue, urlcolor=blue}

\def\dapical{d_\text{apical}}
\def\edp{\epsilon_d - \epsilon_p}
\def\Tcmax{T_\mathrm{c,max}}

\begin{document}

\title{Tuning the charge-transfer energy in hole-doped cuprates}

\author{Chuck-Hou Yee}
\email{chuckyee@physics.rutgers.edu}
\affiliation{Kavli Institute for Theoretical Physics, University of California Santa Barbara, CA 93106, USA}
\author{Gabriel Kotliar}
\affiliation{Department of Physics \& Astronomy, Rutgers University, Piscataway, NJ 08854-8019, USA}

\date{\today}

\begin{abstract}
  Chemical substitution, combined with strain, allows the charge-transfer
  energy in hole-doped cuprates to be broadly tuned.  We theoretically
  characterize the structural and electronic properties of the family of
  compounds $R_2$CuO$_2$S$_2$, constructed by sulfur replacement of the apical
  oxygens and rare earth substitutions in the parent cuprate La$_2$CuO$_4$.
  Additionally, the enthalpies of formation for possible synthesis pathways are
  determined.
\end{abstract}


\maketitle

In weakly correlated systems, analytic and computational tools, such as density
functional theory (DFT), have achieved a level of control where one can
reasonably contemplate materials design.  For recent work in this upcoming
field, see Refs.~\cite{Ceder-MRS, Zunger, fennie, anindya, cein-design}.  The
corresponding methods for strongly-correlated systems are rapidly
advancing~\cite{KotliarRMP, Held, Foulkes-QMC-rmp, Mitas-QMC, gw-review,
  Louie-gw-chapter}. Materials design of correlated systems is a valuable route
to test our physical understanding, and should the materials be synthesized,
provides experimental feedback to accelerate development of theoretical
methods.

The high-temperature cuprate superconductors are classic strongly-correlated
systems which exhibit the highest transition temperatures known. Despite
intense theoretical and experimental study, we have limited first-principles
understanding of the chemical parameters controlling these transition
temperatures.  Based on an \emph{ab initio} study of the
cuprates~\cite{cuprates-charge-transfer}, we proposed that the charge-transfer
energy $\edp$ controls the strength of correlations~\cite{zsa} and therefore
tunes the maximum superconducting transition temperature $\Tcmax$. Starting
with the most correlated cuprate La$_2$CuO$_4$, reducing $\edp$ enhances
$\Tcmax$. In contrast, Sakakibara, \emph{et. al.}, proposed that reducing the
admixture of the Cu $3d_{z^2}$ orbital with the in-plane $3d_{x^2-y^2}$ orbital
enhances $\Tcmax$, a mechanism dubbed ``orbital distillation''~\cite{aoki-prl,
  aoki-prb}. These studies pose a challenge for materials design to test the
theoretical proposals.  

In this work, we design a family of copper oxysulfides to tune the
charge-transfer energy of La$_2$CuO$_4$ (LCO) and test the proposed
theories. We show that substitution of the apical oxygens by sulfur produces
sufficiently large effects on $\edp$ and orbital distillation to warrant
investigation. The compounds are valuable for differentiating the two theories
since, barring competing orders, the proposed mechanisms predict opposite
shifts in $\Tcmax$ in the oxysulfides relative to the LCO baseline. However,
checks on structural stability and the thermodynamics of synthesis pathways
indicate epitaxial stabilization is likely necessary for realization of the
oxysulfides.

\begin{figure}
  \includegraphics[width=\columnwidth]{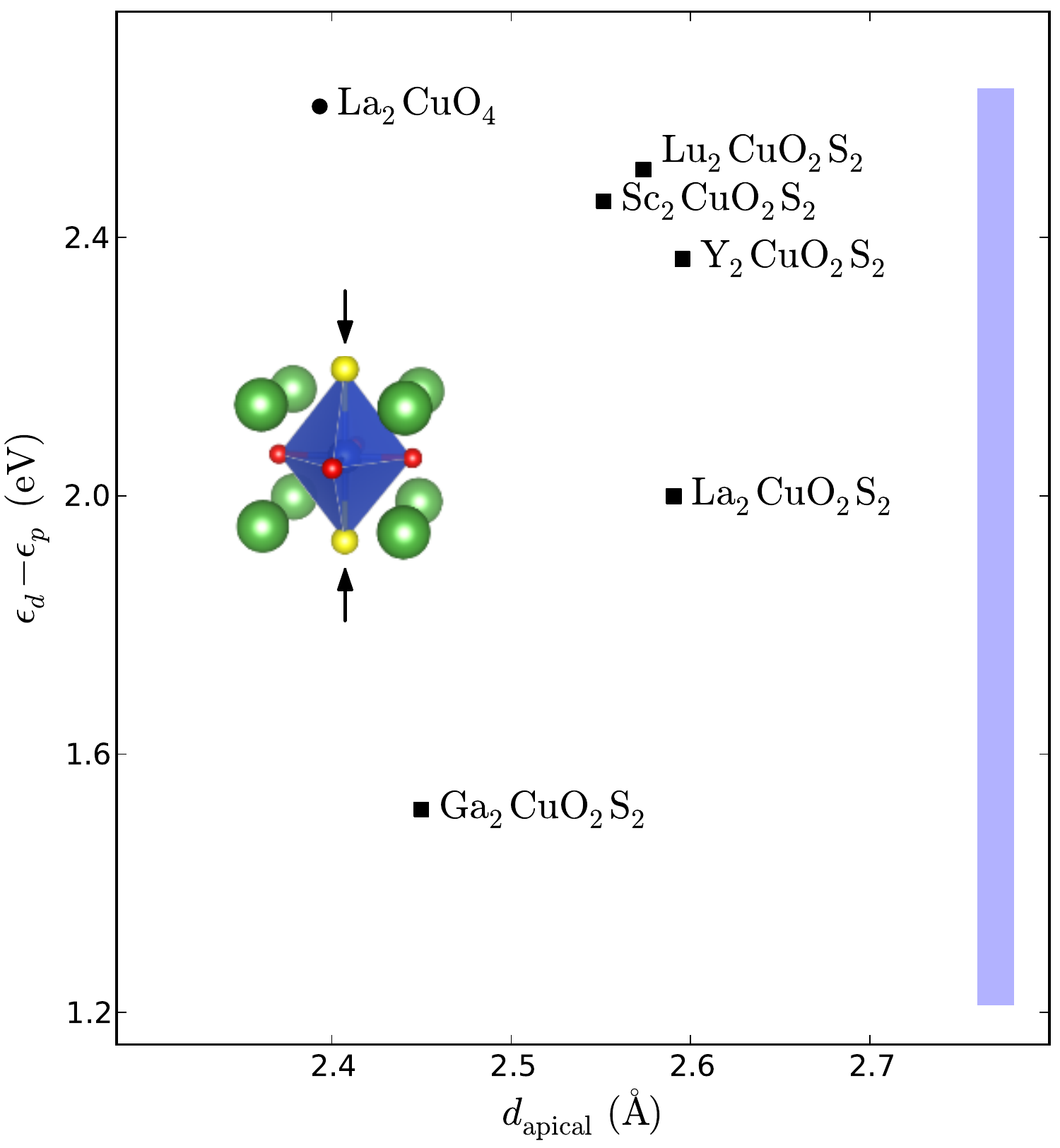}
  \caption{Scatter plot of the charge-transfer energy ($\edp$) and the apical
    atom distance from the CuO$_2$ plane ($\dapical$) for the family of
    hypothetical copper oxysulfides R$_2$CuO$_2$S$_2$ (squares) and the actual
    compound La$_2$CuO$_4$ (circle).  The oxysulfides are formed by the
    substitution of apical oxygens by sulfur (arrows in inset figure of crystal
    structure), and the structure obtained by full relaxation in DFT.  Chemical
    substitution allows us to span nearly the entire range of charge-transfer
    energies (shaded bar) found in the cuprates.}
  \label{fig:edp}
\end{figure}

We choose the $T$-type layered perovskite La$_2$CuO$_4$ (LCO) as our structural
starting point.  This single-layered cuprate is well-studied, supports a broad
range of doping, and exhibits the largest charge-transfer energy of all the
hole-doped cuprates.  Guided by the intuition that larger, more covalent,
apical ions would decrease $\edp$, we replace the apical oxygens in LCO with
sulfur to form the copper oxysulfide La$_2$CuO$_2$S$_2$.  Due to the larger
ionic radii of sulfur relative to oxygen, we expect the LaS charge reservoir
layers to be crowded.  To compensate for this large sulfur in the charge
reservoir layers, we explore the effect substitution of the large La ion with
smaller trivalent ions $R = $Y, Lu, Sc, Ga.  We structurally relax, within a
single unit cell, the entire family of copper oxysulfides $R_2$CuO$_2$S$_2$,
using the Vinet equation of state~\cite{vinet-eos}.  We used the
VASP~\cite{kresse1993, kresse1994a, kresse1996a, kresse1996a} density
functional software package with PAW potentials~\cite{bloechl994, kresse1999}.
Finally, we extract $\edp$ by performing a full-potential calculation on the
relaxed structure using Wien2K~\cite{wien2k}, then
downfolding~\cite{entangled-wannier} to orbitals constructed in the manner
described in Ref.~\cite{haule-ldadmft-prb}.

The charge-transfer energies for the parent compound La$_2$CuO$_4$ and the
family of hypothetical oxysulfides $R_2$CuO$_2$S$_2$ are shown in
Fig.~\ref{fig:edp}.  The substitution of sulfur alone, leaving the La ion
unmodified, decreases the charge-transfer energy by 0.6~eV, a significant
reduction given that $\edp$ spans a range of 1.4~eV across all cuprate
families.  If no competing electronic or structural orders are introduced, we
predict La$_2$CuO$_2$S$_2$ to have a higher superconducting transition
temperature than LCO (which has a $\Tcmax$ of 38~K).  Further substitution of
La by trivalent cations of varying ionic radii allows us to span nearly the
entire range of charge-transfer energies found in the cuprates. We find that
the apical atom distance $\dapical$ is increased by as much as 0.2~\AA in the
oxysulfides, due to the large size of sulfur.

\begin{figure}
  \includegraphics[width=\columnwidth]{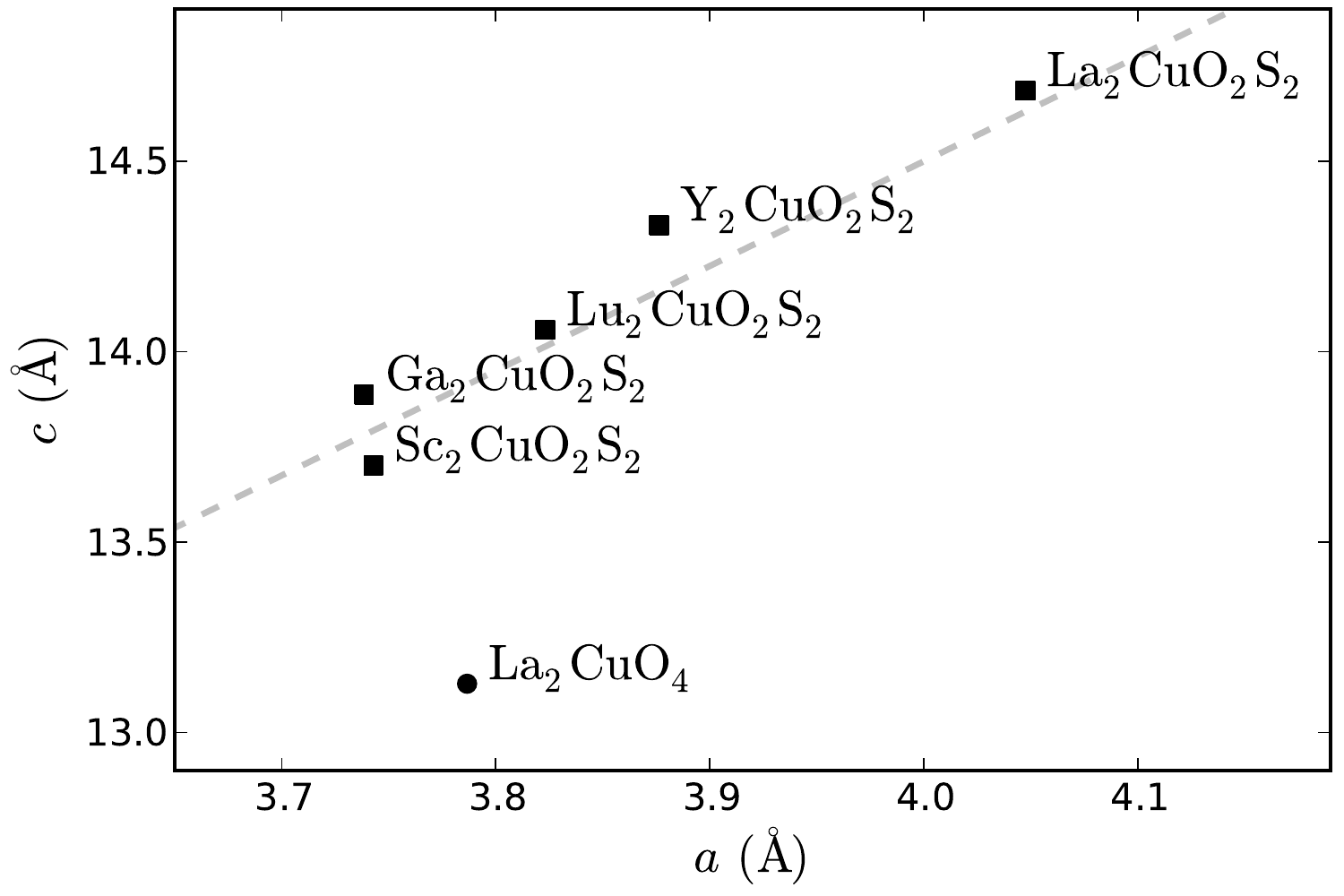}
  \caption{Fully relaxed conventional cell lattice parameters for
    R$_2$CuO$_2$S$_2$ family of compounds (squares) as compared to
    La$_2$CuO$_4$ (dot).  Sulfur substitution increases both the $a$ and $c$
    lattice parameters.  The decrease in the cation's ionic radii from
    (La$\rightarrow$Y$\rightarrow$Lu$\rightarrow$Sc$\rightarrow$Ga) accounts
    for the contraction of $a$ and $c$ within the family of oxysulfides.  The
    dotted line is a guide to the eye.}
  \label{fig:latt-params}
\end{figure}

The structural effects of the sulfur and rare-earth substitutions can be
rationalized entirely due to trends in ionic radii, as shown via a scatter plot
of the $a$- and $c$-axis lattice constants of the proposed compounds
(Fig.~\ref{fig:latt-params}).  The ionic radius of sulfur is greater than that
of oxygen by 0.4~\AA, a significant amount.  Apical sulfur substitution alone
enhances the $c$-axis length by over 1.5~\AA relative to the parent compound
LCO, in agreement with the fact that there are four sulfur layers in the
conventional cell, giving $\Delta c \approx 4 \times 0.4$~\AA.  This chemically
induced strain may be beneficial for superconductivity since it is known
uniaxial compression strongly suppresses $T_\text{c}$ in
La$_2$CuO$_4$~\cite{LSCO-pressure}, a point we will subsequently address.  In
contrast, the in-plane constant exhibits only a slight expansion by 0.2~\AA,
due to the restraining effect of the stiff Cu-O bonds.  The subsequent
rare-earth substitutions in the oxysulfides drive a clear structural trend,
with both lattice parameters contracting in parallel with the decreasing ionic
radius of the rare-earth ion $R$.  The same trend is experimentally observed in
the $T'$-type $R_2$CuO$_4$ family~\cite{LSCO-Tprime}.

\begin{figure}
  \includegraphics[width=\columnwidth]{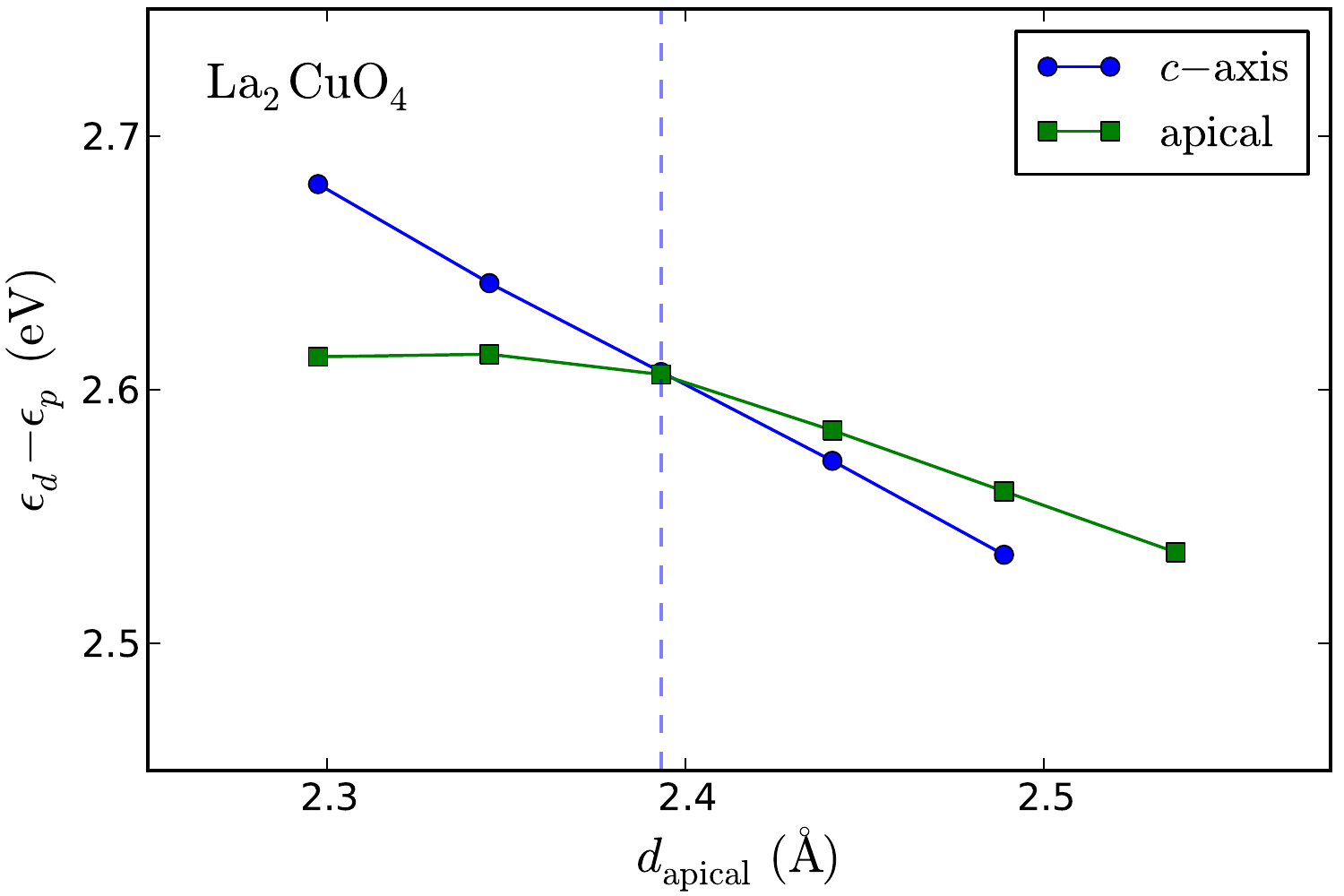}
  \includegraphics[width=\columnwidth]{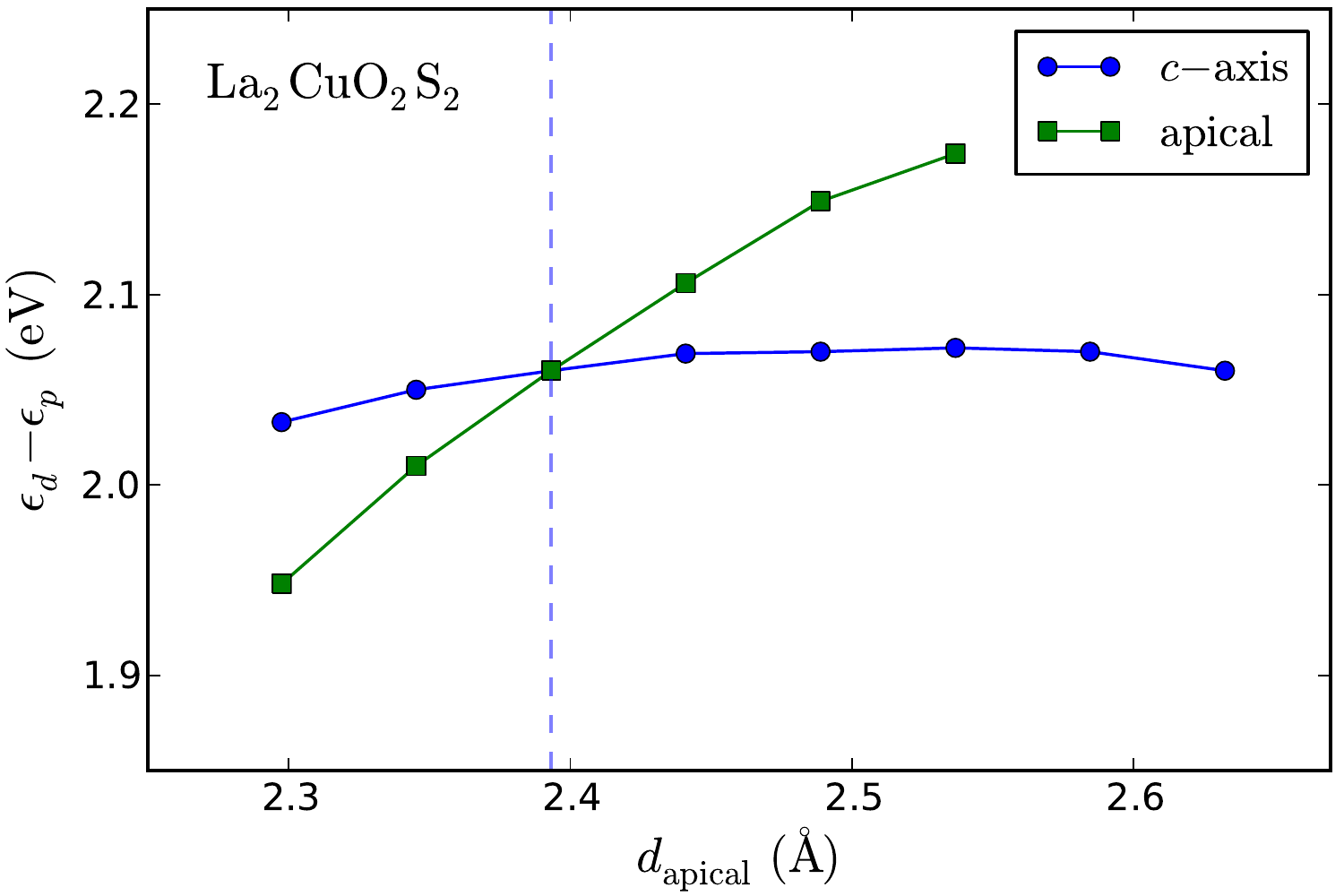}
  \caption{The effect of uniaxial $c$-axis strain on the charge-transfer energy
    (circles) for the actual compound La$_2$CuO$_4$ (top) and the oxysulfide
    La$_2$CuO$_2$S$_2$.  The dotted line shows the equilibrium $\dapical$.  In
    LCO, uniaxial compression increases $\edp$ and consequently suppresses
    $T_\text{c}$, as observed experimentally~\cite{LSCO-pressure}, while we
    predict pressure should hardly affect $T_\text{c}$ in the oxysulfide.  To
    disentangle crystal fields from the Madelung potential, we show the effect
    of tweaking $\dapical$ with all other atoms fixed (squares).  The effects
    due to the Madelung potential, roughly quantified by the difference between
    the two curves, is the same order of magnitude as crystal fields.}
  \label{fig:scan-apical}
\end{figure}

In order to disentangle the structural contributions driving variations in the
charge-transfer energy, we track the change in $\edp$ as we (1) vary the
$c$-axis parameter, modeling uniform uniaxial strain and (2) vary $\dapical$
alone with all other atomic positions fixed.  The difference in the resultant
$\edp$ between these two structural configurations provides a rough proxy for
the magnitude of effects due to the Madelung potential as compared to the local
crystal fields generated in part by the apical atom.  For the parent compound
LCO (Fig.~\ref{fig:scan-apical} top), $c$-axis compression enhances $\edp$.
This increase in correlation strength provides a natural explanation for the
observed suppression of superconductivity with uniaxial
pressure~\cite{LSCO-pressure}.  Modifying $\dapical$ alone has a similar,
albeit smaller, effect.  The difference between the two curves is of the same
magnitude as the total shifts in $\edp$, so we conclude that crystal fields and
Madelung potentials have similar energy scales, making first-principles
calculations necessary for quantitative insight.  In comparison, strain has the
opposite effect in the hypothetical compound La$_2$CuO$_2$S$_2$
(Fig.~\ref{fig:scan-apical} bottom), with $c$-axis compression driving a
reduction in $\edp$.  Thus, $\Tcmax$ should be observed to trend in the
opposite direction as compared to the parent compound.

Finally, we check the structural and thermodynamic stability of the
hypothetical oxysulfides.  Determining the lowest-energy structure of a general
multi-element compound is extremely challenging, and the checks we perform are
by no means exhaustive.  However, guided by knowledge of typical competing
phases and structural distortions, we show the $R_2$CuO$_2$S$_2$ family passes
several basic criteria for stability.

\begin{figure}
  \includegraphics[width=\columnwidth]{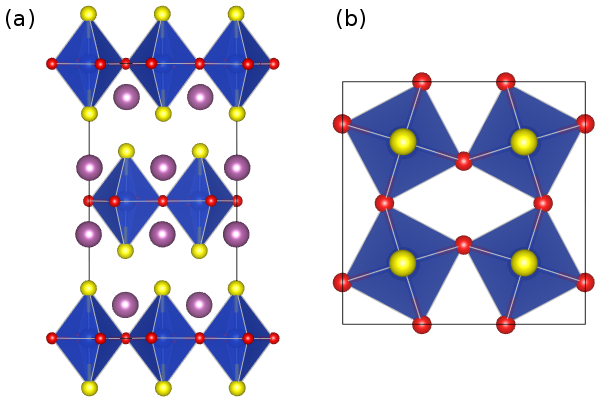}
  \caption{Octahedral rotations in Sc$_2$CuO$_2$S$_2$ as shown by (a) a view
    along the $a$-axis, and (b) a section of the CuO$_2$ plane.  We
    structurally relaxed a 2$\times$2$\times$1 supercell with no symmetry
    constraints.  While there is no out-of-plane buckling of the Cu-O bonds,
    the small Sc ion causes strong in-plane rotations of the octahedral cages
    (denoted $a^0a^0c^-_p$ in Glazer notation).  The compound remains in the
    tetragonal $T$-type phase.}
  \label{fig:scos}
\end{figure}

\begin{table}
  \begin{tabular}{r|r|rl}
    \hline
    $\Delta E$ & $\Delta V$ & \multicolumn{2}{c}{Synthesis pathway} \\
    \hline
    \hline
    141 &  -7.3 &     La$_2$O$_2$S + CuS & $\rightarrow$ La$_2$CuO$_2$S$_2$\\
    223 &  -3.4 &      Y$_2$O$_2$S + CuS & $\rightarrow$  Y$_2$CuO$_2$S$_2$\\
    267 &  -5.0 &     Lu$_2$O$_2$S + CuS & $\rightarrow$ Lu$_2$CuO$_2$S$_2$\\
    356 &  -3.0 &     Sc$_2$O$_2$S + CuS & $\rightarrow$ Sc$_2$CuO$_2$S$_2$\\
    101 &  -4.9 & La$_2$O$_2$S$_2$ + Cu  & $\rightarrow$ La$_2$CuO$_2$S$_2$\\
    \hline
    148 &  -3.3 &      La$_2$O$_3$ + CuS & $\rightarrow$ La$_2$CuO$_3$S \\
    454 &  -0.7 &      Sc$_2$O$_3$ + CuS & $\rightarrow$ Sc$_2$CuO$_3$S \\
     97 &  -4.9 &     La$_2$O$_2$S + CuO & $\rightarrow$ La$_2$CuO$_3$S \\
    269 &   2.8 &     Sc$_2$O$_2$S + CuO & $\rightarrow$ Sc$_2$CuO$_3$S \\
    \hline
     28 &  -5.1 &      La$_2$O$_3$ + CuO & $\rightarrow$ La$_2$CuO$_4$ \\
    \hline
  \end{tabular}
  \caption{Synthesis pathways for various cuprate oxysulfides based on
    substitution of sulfur for both (top block) or only one (middle block) of
    the apical oxygens in $R_2$CuO$_4$.  Energies in kJ/mol and volumes in
    kJ/mol/GPa.  Since the energies of formation ($\Delta E = E_\text{products}
    - E_\text{reactants}$) are positive, none of these pathways appear
    favorable at ambient conditions.  However, high-pressure synthesis will
    help stabilize these pathways, since the majority of volume differentials
    ($\Delta V = V_\text{products} - V_\text{reactants}$) are negative.  We
    benchmark our method against the standard synthesis pathway for
    La$_2$CuO$_4$, shown on the last line.  Surprisingly, $\Delta E$ is
    +28~kJ/mol, so either DFT systemmatically overestimates enthalpies (which
    means the actual enthalpies for our hypothetical compounds are
    \emph{smaller}, in our favor), or we must add a bi-directional uncertainty
    of $\pm 30$~kJ/mol to the computed enthalpies.  Additionally, positional
    entropy of the apical $S$ in the half-substituted $R_2$CuO$_3$S compounds
    should also assist in synthesis.}
  \label{tbl:pathways}
\end{table}

In La$_2$CuO$_4$, when the large La ion is replaced by the smaller Nd and Pr
atoms, the oxygens in the rock-salt LaO layers in the $T$-type structure tend
to be unstable towards rearrangement into the fluorite structure, forming the
$T'$-type structure associated with the electron-doped
cuprates~\cite{T-type-stability}.  We check whether our $T$-type oxysulfides
are locally stable against distortion into the $T'$-type structure by
performing a full structural relaxation in a 2$\times$2$\times$1 supercell with
no symmetry constraints.  We choose Sc$_2$CuO$_2$S$_2$ as a representative
compound, since we believe that the smaller Sc cation will help create room for
the large sulfur anion.  The resultant structure (Fig.~\ref{fig:scos}) shows
strong octahedral rotations along the $c$-axis, denoted $a^0a^0c^-_p$ in Glazer
notation~\cite{Knapp-thesis}.  Perovskite structures often exhibit octahedral
rotations when the Goldschmidt tolerance factor $t = (r_A + r_O)/\sqrt{2}(r_B +
r_O)$ is less than unity as the octahedra rotate to fill the empty space left
by substitution of small rare-earth ions.  However, we find the $T$-type
structure to be locally stable and expect the rotations to be smaller in
magnitude for the larger $R = $La ion.  Additionally, the system remains
tetragonal and shows no out-of-plane buckling of the Cu-O bonds known to be
detrimental to superconductivity~\cite{buckling}.  We note that recent advances
have allowed experimental control over the two competing structures in compound
synthesis~\cite{LSCO-Tprime}, which will help overcome these structural
hurdles.

In addition to local structural stability, we check the thermodynamic stability
of the proposed compounds with respect to competing phases.  We assemble the
computed internal energies of formation $\Delta E = E_\text{products} -
E_\text{reactants}$ in Table~\ref{tbl:pathways} (units of kJ/mol).  That the
quantities are positive imply the that these standard pathways are not
energetically favorable at zero pressure.  However, it is known that many
functional materials are metastable, protected from decay by large energetic
barriers, resulting in technologically useful lifetimes~\cite{Zunger}.  Typical
energetic deficits range from 5 to 30~kJ/mol, when translated to the oxysulfide
compound stoichiometries.  Accompanied by the fact that the volume
differentials $\Delta V = V_\text{products} - V_\text{reactants}$ are negative,
high pressure synthesis may allow the products to form.  Furthermore, we
included in Table~\ref{tbl:pathways} compounds of the form $R_2$CuO$_3$S, where
only half of the apical oxygens are replaced by sulfur.  The configurational
entropy of the sulfur-oxygen arrangement in the charge-reservoir blocks will
further encourage product formation.  These challenges in stability suggest the
oxysulfides could be an ideal system for other materials synthesis techniques
like molecular beam epitaxy.

\begin{figure}
  \includegraphics[width=\columnwidth]{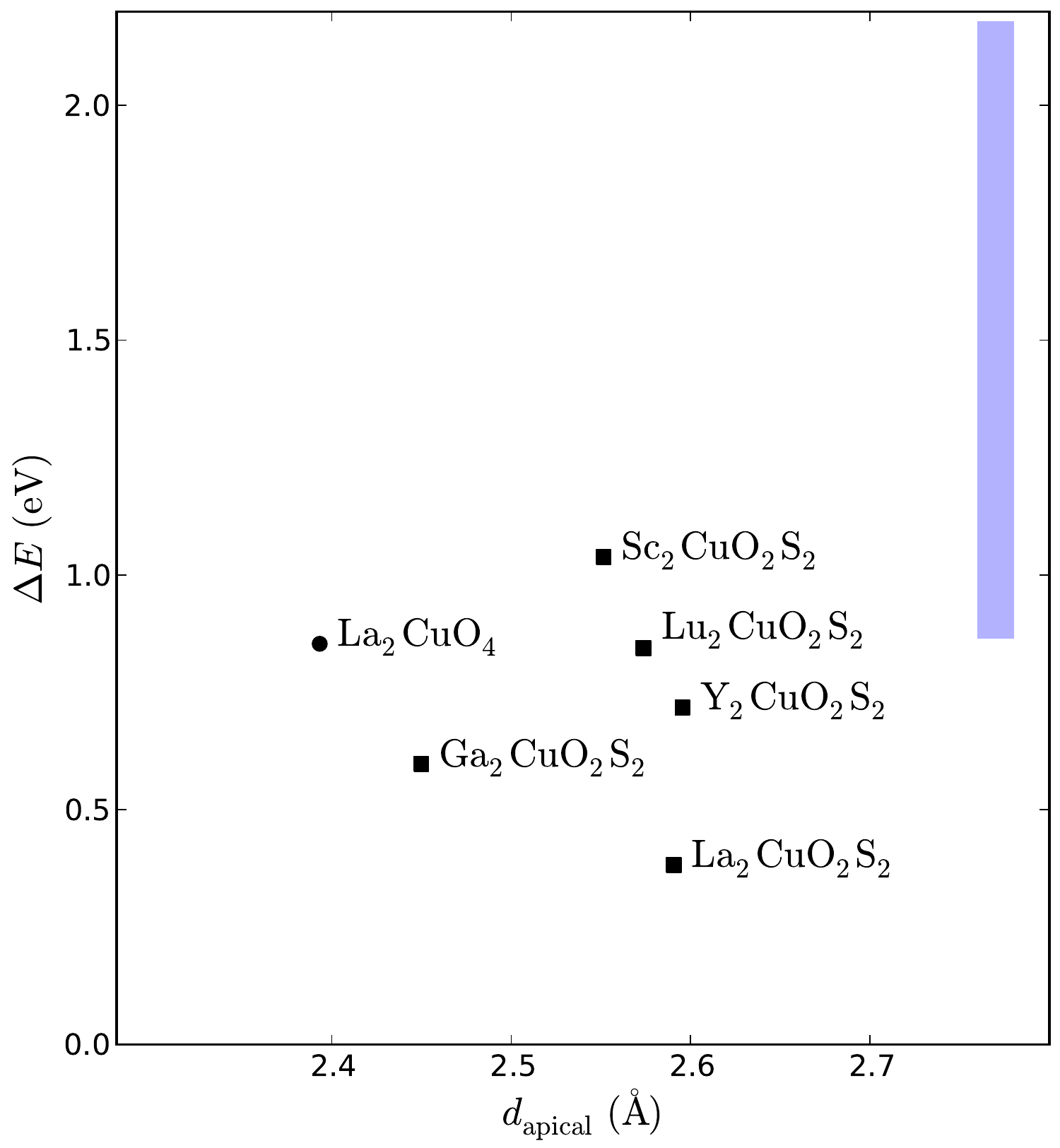}
  \caption{The energy splitting $\Delta E = E_{x^2-y^2} - E_{z^2}$ between the
    onsite energies of the $d_{x^2-y^2}$ and $d_{z^2}$ orbitals among the
    copper oxysulfide family (squares).  For comparison is the actual compound
    La$_2$CuO$_4$ (circle).  The physical range of $\Delta E$ spanned by the
    single-layer cuprates is shown with a shaded bar (from Sakakibara, {\it
      et. al.})~\cite{aoki-prb}.  The $\Delta E$ in the oxysulfides are
    generally smaller, implying stronger mixing of the $d_{z^2}$ orbital with
    the in-plane $d_{x^2-y^2}$ orbital.  Thus, barring other competing orders,
    sulfur substitution should enhance $\Tcmax$ according to the
    charge-transfer theory, while $\Tcmax$ will be suppressed according to the
    two-orbital theory.}
  \label{fig:dz2}
\end{figure}

Turning to the orbital distillation theory, we show in Fig.~\ref{fig:dz2} the
effect of sulfur substitution on the energy splitting $\Delta E =
\epsilon_{x^2-y^2} - \epsilon_{z^2}$, which quantifies the strength of the
$3d_{z^2}$ admixture into the $3d_{x^2-y^2}$ band.  We find that apical
sulfurs, being a larger, more covalent ion than oxygen, generally decreases
$\Delta E$, causing stronger mixing of the $d_{z^2}$ orbital into the
$d_{x^2-y^2}$ band, which suppresses $\Tcmax$ in the two-orbital model.
Barring competing orders, the two proposals predict diverging effects: the
reduction of $\edp$ should enhance $\Tcmax$ in the oxysulfides, while the
increase mixing of $d_{z^2}$ suppress $\Tcmax$.

In conclusion, we have designed via first-principles a family of copper
oxysulfides, based on the well-studied cuprate La$_2$CuO$_4$, which offer broad
tunability in the charge-transfer energy. According to our proposal, the
replacement of oxygen by sulfur in the apical position will enhance the
superconducting $\Tcmax$, while the orbital distillation proposal predicts the
opposite effect. Although we recognize that many other mechanisms can drive
changes in transition temperatures, we nevertheless believe that exploration of
the oxysulfides are a valuable step towards testing the proposed theories.  We
find that molecular beam epitaxy is the preferred synthesis route since epitaxy
will reduce competing orders and tip the energetics away from competing
phases. Natural extensions of this work include generating a detailed \emph{ab
  initio} understanding of the structural handles controlling electronic
parameters and pursuing similar chemical engineering of crystal field levels in
the Bi, Tl and Hg-based cuprate families.

C.Y. acknowledges support from MRL grant DMR-11-21053 and KITP grant
PHY-11-25915. G.K acknowledges support from NSF grant No. DMR-0906943.

\bibliography{tuning-charge-transfer,vaspref}

\begin{thebibliography}{29}%
\makeatletter
\providecommand \@ifxundefined [1]{%
 \@ifx{#1\undefined}
}%
\providecommand \@ifnum [1]{%
 \ifnum #1\expandafter \@firstoftwo
 \else \expandafter \@secondoftwo
 \fi
}%
\providecommand \@ifx [1]{%
 \ifx #1\expandafter \@firstoftwo
 \else \expandafter \@secondoftwo
 \fi
}%
\providecommand \natexlab [1]{#1}%
\providecommand \enquote  [1]{``#1''}%
\providecommand \bibnamefont  [1]{#1}%
\providecommand \bibfnamefont [1]{#1}%
\providecommand \citenamefont [1]{#1}%
\providecommand \href@noop [0]{\@secondoftwo}%
\providecommand \href [0]{\begingroup \@sanitize@url \@href}%
\providecommand \@href[1]{\@@startlink{#1}\@@href}%
\providecommand \@@href[1]{\endgroup#1\@@endlink}%
\providecommand \@sanitize@url [0]{\catcode `\\12\catcode `\$12\catcode
  `\&12\catcode `\#12\catcode `\^12\catcode `\_12\catcode `\%12\relax}%
\providecommand \@@startlink[1]{}%
\providecommand \@@endlink[0]{}%
\providecommand \url  [0]{\begingroup\@sanitize@url \@url }%
\providecommand \@url [1]{\endgroup\@href {#1}{\urlprefix }}%
\providecommand \urlprefix  [0]{URL }%
\providecommand \Eprint [0]{\href }%
\providecommand \doibase [0]{http://dx.doi.org/}%
\providecommand \selectlanguage [0]{\@gobble}%
\providecommand \bibinfo  [0]{\@secondoftwo}%
\providecommand \bibfield  [0]{\@secondoftwo}%
\providecommand \translation [1]{[#1]}%
\providecommand \BibitemOpen [0]{}%
\providecommand \bibitemStop [0]{}%
\providecommand \bibitemNoStop [0]{.\EOS\space}%
\providecommand \EOS [0]{\spacefactor3000\relax}%
\providecommand \BibitemShut  [1]{\csname bibitem#1\endcsname}%
\let\auto@bib@innerbib\@empty
\bibitem [{\citenamefont {Hafner}\ \emph {et~al.}(2006)\citenamefont {Hafner},
  \citenamefont {Wolverton},\ and\ \citenamefont {Ceder}}]{Ceder-MRS}%
  \BibitemOpen
  \bibfield  {author} {\bibinfo {author} {\bibfnamefont {J.}~\bibnamefont
  {Hafner}}, \bibinfo {author} {\bibfnamefont {C.}~\bibnamefont {Wolverton}}, \
  and\ \bibinfo {author} {\bibfnamefont {G.}~\bibnamefont {Ceder}},\ }\href
  {http://dx.doi.org/10.1557/mrs2006.174} {\bibfield  {journal} {\bibinfo
  {journal} {MRS Bulletin}\ }\textbf {\bibinfo {volume} {31}},\ \bibinfo
  {pages} {659} (\bibinfo {year} {2006})}\BibitemShut {NoStop}%
\bibitem [{\citenamefont {Zhang}\ \emph {et~al.}(2012)\citenamefont {Zhang},
  \citenamefont {Yu}, \citenamefont {Zakutayev},\ and\ \citenamefont
  {Zunger}}]{Zunger}%
  \BibitemOpen
  \bibfield  {author} {\bibinfo {author} {\bibfnamefont {X.}~\bibnamefont
  {Zhang}}, \bibinfo {author} {\bibfnamefont {L.}~\bibnamefont {Yu}}, \bibinfo
  {author} {\bibfnamefont {A.}~\bibnamefont {Zakutayev}}, \ and\ \bibinfo
  {author} {\bibfnamefont {A.}~\bibnamefont {Zunger}},\ }\href {\doibase
  10.1002/adfm.201102546} {\bibfield  {journal} {\bibinfo  {journal} {Advanced
  Functional Materials}\ }\textbf {\bibinfo {volume} {22}},\ \bibinfo {pages}
  {1425} (\bibinfo {year} {2012})}\BibitemShut {NoStop}%
\bibitem [{\citenamefont {Fennie}(2008)}]{fennie}%
  \BibitemOpen
  \bibfield  {author} {\bibinfo {author} {\bibfnamefont {C.~J.}\ \bibnamefont
  {Fennie}},\ }\href {\doibase 10.1103/PhysRevLett.100.167203} {\bibfield
  {journal} {\bibinfo  {journal} {Phys. Rev. Lett.}\ }\textbf {\bibinfo
  {volume} {100}},\ \bibinfo {pages} {167203} (\bibinfo {year}
  {2008})}\BibitemShut {NoStop}%
\bibitem [{\citenamefont {Roy}\ \emph {et~al.}(2012)\citenamefont {Roy},
  \citenamefont {Bennett}, \citenamefont {Rabe},\ and\ \citenamefont
  {Vanderbilt}}]{anindya}%
  \BibitemOpen
  \bibfield  {author} {\bibinfo {author} {\bibfnamefont {A.}~\bibnamefont
  {Roy}}, \bibinfo {author} {\bibfnamefont {J.~W.}\ \bibnamefont {Bennett}},
  \bibinfo {author} {\bibfnamefont {K.~M.}\ \bibnamefont {Rabe}}, \ and\
  \bibinfo {author} {\bibfnamefont {D.}~\bibnamefont {Vanderbilt}},\ }\href
  {\doibase 10.1103/PhysRevLett.109.037602} {\bibfield  {journal} {\bibinfo
  {journal} {Phys. Rev. Lett.}\ }\textbf {\bibinfo {volume} {109}},\ \bibinfo
  {pages} {037602} (\bibinfo {year} {2012})}\BibitemShut {NoStop}%
\bibitem [{\citenamefont {{Fredeman}}\ \emph {et~al.}(2011)\citenamefont
  {{Fredeman}}, \citenamefont {{Tobash}}, \citenamefont {{Torrez}},
  \citenamefont {{Thompson}}, \citenamefont {{Bauer}}, \citenamefont
  {{Ronning}}, \citenamefont {{Tipton}}, \citenamefont {{Rudin}},\ and\
  \citenamefont {{Hennig}}}]{cein-design}%
  \BibitemOpen
  \bibfield  {author} {\bibinfo {author} {\bibfnamefont {D.~J.}\ \bibnamefont
  {{Fredeman}}}, \bibinfo {author} {\bibfnamefont {P.~H.}\ \bibnamefont
  {{Tobash}}}, \bibinfo {author} {\bibfnamefont {M.~A.}\ \bibnamefont
  {{Torrez}}}, \bibinfo {author} {\bibfnamefont {J.~D.}\ \bibnamefont
  {{Thompson}}}, \bibinfo {author} {\bibfnamefont {E.~D.}\ \bibnamefont
  {{Bauer}}}, \bibinfo {author} {\bibfnamefont {F.}~\bibnamefont {{Ronning}}},
  \bibinfo {author} {\bibfnamefont {W.~W.}\ \bibnamefont {{Tipton}}}, \bibinfo
  {author} {\bibfnamefont {S.~P.}\ \bibnamefont {{Rudin}}}, \ and\ \bibinfo
  {author} {\bibfnamefont {R.~G.}\ \bibnamefont {{Hennig}}},\ }\href {\doibase
  10.1103/PhysRevB.83.224102} {\bibfield  {journal} {\bibinfo  {journal}
  {\prb}\ }\textbf {\bibinfo {volume} {83}},\ \bibinfo {eid} {224102} (\bibinfo
  {year} {2011})},\ \Eprint {http://arxiv.org/abs/1103.2259} {arXiv:1103.2259
  [cond-mat.mtrl-sci]} \BibitemShut {NoStop}%
\bibitem [{\citenamefont {Kotliar}\ \emph {et~al.}(2006)\citenamefont
  {Kotliar}, \citenamefont {Savrasov}, \citenamefont {Haule}, \citenamefont
  {Oudovenko}, \citenamefont {Parcollet},\ and\ \citenamefont
  {Marianetti}}]{KotliarRMP}%
  \BibitemOpen
  \bibfield  {author} {\bibinfo {author} {\bibfnamefont {G.}~\bibnamefont
  {Kotliar}}, \bibinfo {author} {\bibfnamefont {S.~Y.}\ \bibnamefont
  {Savrasov}}, \bibinfo {author} {\bibfnamefont {K.}~\bibnamefont {Haule}},
  \bibinfo {author} {\bibfnamefont {V.~S.}\ \bibnamefont {Oudovenko}}, \bibinfo
  {author} {\bibfnamefont {O.}~\bibnamefont {Parcollet}}, \ and\ \bibinfo
  {author} {\bibfnamefont {C.~A.}\ \bibnamefont {Marianetti}},\ }\href
  {\doibase 10.1103/RevModPhys.78.865} {\bibfield  {journal} {\bibinfo
  {journal} {Rev. Mod. Phys.}\ }\textbf {\bibinfo {volume} {78}},\ \bibinfo
  {eid} {865} (\bibinfo {year} {2006})}\BibitemShut {NoStop}%
\bibitem [{\citenamefont {Held}(2007)}]{Held}%
  \BibitemOpen
  \bibfield  {author} {\bibinfo {author} {\bibfnamefont {K.}~\bibnamefont
  {Held}},\ }\href@noop {} {\bibfield  {journal} {\bibinfo  {journal} {Advances
  in Physics}\ }\textbf {\bibinfo {volume} {56}},\ \bibinfo {pages} {829}
  (\bibinfo {year} {2007})}\BibitemShut {NoStop}%
\bibitem [{\citenamefont {Foulkes}\ \emph {et~al.}(2001)\citenamefont
  {Foulkes}, \citenamefont {Mitas}, \citenamefont {Needs},\ and\ \citenamefont
  {Rajagopal}}]{Foulkes-QMC-rmp}%
  \BibitemOpen
  \bibfield  {author} {\bibinfo {author} {\bibfnamefont {W.~M.~C.}\
  \bibnamefont {Foulkes}}, \bibinfo {author} {\bibfnamefont {L.}~\bibnamefont
  {Mitas}}, \bibinfo {author} {\bibfnamefont {R.~J.}\ \bibnamefont {Needs}}, \
  and\ \bibinfo {author} {\bibfnamefont {G.}~\bibnamefont {Rajagopal}},\ }\href
  {\doibase 10.1103/RevModPhys.73.33} {\bibfield  {journal} {\bibinfo
  {journal} {Rev. Mod. Phys.}\ }\textbf {\bibinfo {volume} {73}},\ \bibinfo
  {pages} {33} (\bibinfo {year} {2001})}\BibitemShut {NoStop}%
\bibitem [{\citenamefont {Bajdich}\ and\ \citenamefont
  {Mitas}(2009)}]{Mitas-QMC}%
  \BibitemOpen
  \bibfield  {author} {\bibinfo {author} {\bibfnamefont {M.}~\bibnamefont
  {Bajdich}}\ and\ \bibinfo {author} {\bibfnamefont {L.}~\bibnamefont
  {Mitas}},\ }\href@noop {} {\bibfield  {journal} {\bibinfo  {journal} {Acta
  Physica Slovaca}\ }\textbf {\bibinfo {volume} {59}},\ \bibinfo {pages} {81}
  (\bibinfo {year} {2009})}\BibitemShut {NoStop}%
\bibitem [{\citenamefont {Onida}\ \emph {et~al.}(2002)\citenamefont {Onida},
  \citenamefont {Reining},\ and\ \citenamefont {Rubio}}]{gw-review}%
  \BibitemOpen
  \bibfield  {author} {\bibinfo {author} {\bibfnamefont {G.}~\bibnamefont
  {Onida}}, \bibinfo {author} {\bibfnamefont {L.}~\bibnamefont {Reining}}, \
  and\ \bibinfo {author} {\bibfnamefont {A.}~\bibnamefont {Rubio}},\ }\href
  {\doibase 10.1103/RevModPhys.74.601} {\bibfield  {journal} {\bibinfo
  {journal} {Rev. Mod. Phys.}\ }\textbf {\bibinfo {volume} {74}},\ \bibinfo
  {pages} {601} (\bibinfo {year} {2002})}\BibitemShut {NoStop}%
\bibitem [{\citenamefont {Louie}(2006)}]{Louie-gw-chapter}%
  \BibitemOpen
  \bibfield  {author} {\bibinfo {author} {\bibfnamefont {S.}~\bibnamefont
  {Louie}},\ }in\ \href {\doibase 10.1016/S1572-0934(06)02002-6} {\emph
  {\bibinfo {booktitle} {Conceptual Foundations of Materials A Standard Model
  for Ground- and Excited-State Properties}}},\ \bibinfo {series} {Contemporary
  Concepts of Condensed Matter Science}, Vol.~\bibinfo {volume} {2},\ \bibinfo
  {editor} {edited by\ \bibinfo {editor} {\bibfnamefont {S.~G.}\ \bibnamefont
  {Louie}}\ and\ \bibinfo {editor} {\bibfnamefont {M.~L.}\ \bibnamefont
  {Cohen}}}\ (\bibinfo  {publisher} {Elsevier},\ \bibinfo {year} {2006})\ pp.\
  \bibinfo {pages} {9 -- 53}\BibitemShut {NoStop}%
\bibitem [{\citenamefont {Weber}\ \emph {et~al.}(2012)\citenamefont {Weber},
  \citenamefont {Yee}, \citenamefont {Haule},\ and\ \citenamefont
  {Kotliar}}]{cuprates-charge-transfer}%
  \BibitemOpen
  \bibfield  {author} {\bibinfo {author} {\bibfnamefont {C.}~\bibnamefont
  {Weber}}, \bibinfo {author} {\bibfnamefont {C.}~\bibnamefont {Yee}}, \bibinfo
  {author} {\bibfnamefont {K.}~\bibnamefont {Haule}}, \ and\ \bibinfo {author}
  {\bibfnamefont {G.}~\bibnamefont {Kotliar}},\ }\href
  {http://stacks.iop.org/0295-5075/100/i=3/a=37001} {\bibfield  {journal}
  {\bibinfo  {journal} {EPL (Europhysics Letters)}\ }\textbf {\bibinfo {volume}
  {100}},\ \bibinfo {pages} {37001} (\bibinfo {year} {2012})}\BibitemShut
  {NoStop}%
\bibitem [{\citenamefont {Zaanen}\ \emph {et~al.}(1985)\citenamefont {Zaanen},
  \citenamefont {Sawatzky},\ and\ \citenamefont {Allen}}]{zsa}%
  \BibitemOpen
  \bibfield  {author} {\bibinfo {author} {\bibfnamefont {J.}~\bibnamefont
  {Zaanen}}, \bibinfo {author} {\bibfnamefont {G.~A.}\ \bibnamefont
  {Sawatzky}}, \ and\ \bibinfo {author} {\bibfnamefont {J.~W.}\ \bibnamefont
  {Allen}},\ }\href {\doibase 10.1103/PhysRevLett.55.418} {\bibfield  {journal}
  {\bibinfo  {journal} {Phys. Rev. Lett.}\ }\textbf {\bibinfo {volume} {55}},\
  \bibinfo {pages} {418} (\bibinfo {year} {1985})}\BibitemShut {NoStop}%
\bibitem [{\citenamefont {Sakakibara}\ \emph {et~al.}(2010)\citenamefont
  {Sakakibara}, \citenamefont {Usui}, \citenamefont {Kuroki}, \citenamefont
  {Arita},\ and\ \citenamefont {Aoki}}]{aoki-prl}%
  \BibitemOpen
  \bibfield  {author} {\bibinfo {author} {\bibfnamefont {H.}~\bibnamefont
  {Sakakibara}}, \bibinfo {author} {\bibfnamefont {H.}~\bibnamefont {Usui}},
  \bibinfo {author} {\bibfnamefont {K.}~\bibnamefont {Kuroki}}, \bibinfo
  {author} {\bibfnamefont {R.}~\bibnamefont {Arita}}, \ and\ \bibinfo {author}
  {\bibfnamefont {H.}~\bibnamefont {Aoki}},\ }\href {\doibase
  10.1103/PhysRevLett.105.057003} {\bibfield  {journal} {\bibinfo  {journal}
  {Phys. Rev. Lett.}\ }\textbf {\bibinfo {volume} {105}},\ \bibinfo {pages}
  {057003} (\bibinfo {year} {2010})}\BibitemShut {NoStop}%
\bibitem [{\citenamefont {Sakakibara}\ \emph {et~al.}(2012)\citenamefont
  {Sakakibara}, \citenamefont {Usui}, \citenamefont {Kuroki}, \citenamefont
  {Arita},\ and\ \citenamefont {Aoki}}]{aoki-prb}%
  \BibitemOpen
  \bibfield  {author} {\bibinfo {author} {\bibfnamefont {H.}~\bibnamefont
  {Sakakibara}}, \bibinfo {author} {\bibfnamefont {H.}~\bibnamefont {Usui}},
  \bibinfo {author} {\bibfnamefont {K.}~\bibnamefont {Kuroki}}, \bibinfo
  {author} {\bibfnamefont {R.}~\bibnamefont {Arita}}, \ and\ \bibinfo {author}
  {\bibfnamefont {H.}~\bibnamefont {Aoki}},\ }\href {\doibase
  10.1103/PhysRevB.85.064501} {\bibfield  {journal} {\bibinfo  {journal} {Phys.
  Rev. B}\ }\textbf {\bibinfo {volume} {85}},\ \bibinfo {pages} {064501}
  (\bibinfo {year} {2012})}\BibitemShut {NoStop}%
\bibitem [{\citenamefont {Vinet}\ \emph {et~al.}(1989)\citenamefont {Vinet},
  \citenamefont {Rose}, \citenamefont {Ferrante},\ and\ \citenamefont
  {Smith}}]{vinet-eos}%
  \BibitemOpen
  \bibfield  {author} {\bibinfo {author} {\bibfnamefont {P.}~\bibnamefont
  {Vinet}}, \bibinfo {author} {\bibfnamefont {J.~H.}\ \bibnamefont {Rose}},
  \bibinfo {author} {\bibfnamefont {J.}~\bibnamefont {Ferrante}}, \ and\
  \bibinfo {author} {\bibfnamefont {J.~R.}\ \bibnamefont {Smith}},\ }\href
  {http://stacks.iop.org/0953-8984/1/i=11/a=002} {\bibfield  {journal}
  {\bibinfo  {journal} {Journal of Physics: Condensed Matter}\ }\textbf
  {\bibinfo {volume} {1}},\ \bibinfo {pages} {1941} (\bibinfo {year}
  {1989})}\BibitemShut {NoStop}%
\bibitem [{\citenamefont {Kresse}\ and\ \citenamefont
  {Hafner}(1993)}]{kresse1993}%
  \BibitemOpen
  \bibfield  {author} {\bibinfo {author} {\bibfnamefont {G.}~\bibnamefont
  {Kresse}}\ and\ \bibinfo {author} {\bibfnamefont {J.}~\bibnamefont
  {Hafner}},\ }\href@noop {} {\bibfield  {journal} {\bibinfo  {journal} {Phys.
  Rev. B}\ }\textbf {\bibinfo {volume} {47}},\ \bibinfo {pages} {558} (\bibinfo
  {year} {1993})}\BibitemShut {NoStop}%
\bibitem [{\citenamefont {Kresse}\ and\ \citenamefont
  {Hafner}(1994)}]{kresse1994a}%
  \BibitemOpen
  \bibfield  {author} {\bibinfo {author} {\bibfnamefont {G.}~\bibnamefont
  {Kresse}}\ and\ \bibinfo {author} {\bibfnamefont {J.}~\bibnamefont
  {Hafner}},\ }\href@noop {} {\bibfield  {journal} {\bibinfo  {journal} {Phys.
  Rev. B}\ }\textbf {\bibinfo {volume} {49}},\ \bibinfo {pages} {14251}
  (\bibinfo {year} {1994})}\BibitemShut {NoStop}%
\bibitem [{\citenamefont {Kresse}\ and\ \citenamefont
  {Furthmüller}(1996)}]{kresse1996a}%
  \BibitemOpen
  \bibfield  {author} {\bibinfo {author} {\bibfnamefont {G.}~\bibnamefont
  {Kresse}}\ and\ \bibinfo {author} {\bibfnamefont {J.}~\bibnamefont
  {Furthmüller}},\ }\href@noop {} {\bibfield  {journal} {\bibinfo  {journal}
  {Comput. Mat. Sci.}\ }\textbf {\bibinfo {volume} {6}},\ \bibinfo {pages} {15}
  (\bibinfo {year} {1996})}\BibitemShut {NoStop}%
\bibitem [{\citenamefont {Bl\"{o}chl}(1994)}]{bloechl994}%
  \BibitemOpen
  \bibfield  {author} {\bibinfo {author} {\bibfnamefont {P.~E.}\ \bibnamefont
  {Bl\"{o}chl}},\ }\href@noop {} {\bibfield  {journal} {\bibinfo  {journal}
  {Phys. Rev. B}\ }\textbf {\bibinfo {volume} {50}},\ \bibinfo {pages} {17953}
  (\bibinfo {year} {1994})}\BibitemShut {NoStop}%
\bibitem [{\citenamefont {Kresse}\ and\ \citenamefont
  {Joubert}(1999)}]{kresse1999}%
  \BibitemOpen
  \bibfield  {author} {\bibinfo {author} {\bibfnamefont {G.}~\bibnamefont
  {Kresse}}\ and\ \bibinfo {author} {\bibfnamefont {D.}~\bibnamefont
  {Joubert}},\ }\href@noop {} {\bibfield  {journal} {\bibinfo  {journal} {Phys.
  Rev. B}\ }\textbf {\bibinfo {volume} {59}},\ \bibinfo {pages} {1758}
  (\bibinfo {year} {1999})}\BibitemShut {NoStop}%
\bibitem [{\citenamefont {Blaha}\ \emph {et~al.}(2001)\citenamefont {Blaha},
  \citenamefont {Schwarz}, \citenamefont {Madsen}, \citenamefont {Kvasnicka},\
  and\ \citenamefont {Luitz}}]{wien2k}%
  \BibitemOpen
  \bibfield  {author} {\bibinfo {author} {\bibfnamefont {P.}~\bibnamefont
  {Blaha}}, \bibinfo {author} {\bibfnamefont {K.}~\bibnamefont {Schwarz}},
  \bibinfo {author} {\bibfnamefont {G.~K.~H.}\ \bibnamefont {Madsen}}, \bibinfo
  {author} {\bibfnamefont {D.}~\bibnamefont {Kvasnicka}}, \ and\ \bibinfo
  {author} {\bibfnamefont {J.}~\bibnamefont {Luitz}},\ }\href@noop {} {\emph
  {\bibinfo {title} {{WIEN2K}, {A}n {A}ugmented {P}lane {W}ave + {L}ocal
  {O}rbitals {P}rogram for {C}alculating {C}rystal {P}roperties}}}\ (\bibinfo
  {publisher} {{K}arlheinz Schwarz, Techn. Universit\"{a}t Wien, Austria},\
  \bibinfo {year} {2001})\BibitemShut {NoStop}%
\bibitem [{\citenamefont {Souza}\ \emph {et~al.}(2001)\citenamefont {Souza},
  \citenamefont {Marzari},\ and\ \citenamefont
  {Vanderbilt}}]{entangled-wannier}%
  \BibitemOpen
  \bibfield  {author} {\bibinfo {author} {\bibfnamefont {I.}~\bibnamefont
  {Souza}}, \bibinfo {author} {\bibfnamefont {N.}~\bibnamefont {Marzari}}, \
  and\ \bibinfo {author} {\bibfnamefont {D.}~\bibnamefont {Vanderbilt}},\
  }\href {\doibase 10.1103/PhysRevB.65.035109} {\bibfield  {journal} {\bibinfo
  {journal} {Phys. Rev. B}\ }\textbf {\bibinfo {volume} {65}},\ \bibinfo
  {pages} {035109} (\bibinfo {year} {2001})}\BibitemShut {NoStop}%
\bibitem [{\citenamefont {Haule}\ \emph {et~al.}(2010)\citenamefont {Haule},
  \citenamefont {Yee},\ and\ \citenamefont {Kim}}]{haule-ldadmft-prb}%
  \BibitemOpen
  \bibfield  {author} {\bibinfo {author} {\bibfnamefont {K.}~\bibnamefont
  {Haule}}, \bibinfo {author} {\bibfnamefont {C.-H.}\ \bibnamefont {Yee}}, \
  and\ \bibinfo {author} {\bibfnamefont {K.}~\bibnamefont {Kim}},\ }\href
  {\doibase 10.1103/PhysRevB.81.195107} {\bibfield  {journal} {\bibinfo
  {journal} {Phys. Rev. B}\ }\textbf {\bibinfo {volume} {81}},\ \bibinfo
  {pages} {195107} (\bibinfo {year} {2010})}\BibitemShut {NoStop}%
\bibitem [{\citenamefont {Locquet}\ \emph {et~al.}(1998)\citenamefont
  {Locquet}, \citenamefont {Perret}, \citenamefont {Fompeyrine}, \citenamefont
  {Mächler}, \citenamefont {Seo},\ and\ \citenamefont
  {Tendeloo}}]{LSCO-pressure}%
  \BibitemOpen
  \bibfield  {author} {\bibinfo {author} {\bibfnamefont {J.-P.}\ \bibnamefont
  {Locquet}}, \bibinfo {author} {\bibfnamefont {J.}~\bibnamefont {Perret}},
  \bibinfo {author} {\bibfnamefont {J.}~\bibnamefont {Fompeyrine}}, \bibinfo
  {author} {\bibfnamefont {E.}~\bibnamefont {Mächler}}, \bibinfo {author}
  {\bibfnamefont {J.~W.}\ \bibnamefont {Seo}}, \ and\ \bibinfo {author}
  {\bibfnamefont {G.~V.}\ \bibnamefont {Tendeloo}},\ }\href@noop {} {\bibfield
  {journal} {\bibinfo  {journal} {Nature}\ }\textbf {\bibinfo {volume} {394}},\
  \bibinfo {pages} {453} (\bibinfo {year} {1998})}\BibitemShut {NoStop}%
\bibitem [{\citenamefont {Hord}\ \emph {et~al.}(2010)\citenamefont {Hord},
  \citenamefont {Luetkens}, \citenamefont {Pascua}, \citenamefont {Buckow},
  \citenamefont {Hofmann}, \citenamefont {Krockenberger}, \citenamefont
  {Kurian}, \citenamefont {Maeter}, \citenamefont {Klauss}, \citenamefont
  {Pomjakushin}, \citenamefont {Suter}, \citenamefont {Albert},\ and\
  \citenamefont {Alff}}]{LSCO-Tprime}%
  \BibitemOpen
  \bibfield  {author} {\bibinfo {author} {\bibfnamefont {R.}~\bibnamefont
  {Hord}}, \bibinfo {author} {\bibfnamefont {H.}~\bibnamefont {Luetkens}},
  \bibinfo {author} {\bibfnamefont {G.}~\bibnamefont {Pascua}}, \bibinfo
  {author} {\bibfnamefont {A.}~\bibnamefont {Buckow}}, \bibinfo {author}
  {\bibfnamefont {K.}~\bibnamefont {Hofmann}}, \bibinfo {author} {\bibfnamefont
  {Y.}~\bibnamefont {Krockenberger}}, \bibinfo {author} {\bibfnamefont
  {J.}~\bibnamefont {Kurian}}, \bibinfo {author} {\bibfnamefont
  {H.}~\bibnamefont {Maeter}}, \bibinfo {author} {\bibfnamefont {H.-H.}\
  \bibnamefont {Klauss}}, \bibinfo {author} {\bibfnamefont {V.}~\bibnamefont
  {Pomjakushin}}, \bibinfo {author} {\bibfnamefont {A.}~\bibnamefont {Suter}},
  \bibinfo {author} {\bibfnamefont {B.}~\bibnamefont {Albert}}, \ and\ \bibinfo
  {author} {\bibfnamefont {L.}~\bibnamefont {Alff}},\ }\href {\doibase
  10.1103/PhysRevB.82.180508} {\bibfield  {journal} {\bibinfo  {journal} {Phys.
  Rev. B}\ }\textbf {\bibinfo {volume} {82}},\ \bibinfo {pages} {180508}
  (\bibinfo {year} {2010})}\BibitemShut {NoStop}%
\bibitem [{\citenamefont {Tsukada}\ \emph {et~al.}(2005)\citenamefont
  {Tsukada}, \citenamefont {Krockenberger}, \citenamefont {Noda}, \citenamefont
  {Yamamoto}, \citenamefont {Manske}, \citenamefont {Alff},\ and\ \citenamefont
  {Naito}}]{T-type-stability}%
  \BibitemOpen
  \bibfield  {author} {\bibinfo {author} {\bibfnamefont {A.}~\bibnamefont
  {Tsukada}}, \bibinfo {author} {\bibfnamefont {Y.}~\bibnamefont
  {Krockenberger}}, \bibinfo {author} {\bibfnamefont {M.}~\bibnamefont {Noda}},
  \bibinfo {author} {\bibfnamefont {H.}~\bibnamefont {Yamamoto}}, \bibinfo
  {author} {\bibfnamefont {D.}~\bibnamefont {Manske}}, \bibinfo {author}
  {\bibfnamefont {L.}~\bibnamefont {Alff}}, \ and\ \bibinfo {author}
  {\bibfnamefont {M.}~\bibnamefont {Naito}},\ }\href {\doibase
  10.1016/j.ssc.2004.12.011} {\bibfield  {journal} {\bibinfo  {journal} {Solid
  State Communications}\ }\textbf {\bibinfo {volume} {133}},\ \bibinfo {pages}
  {427 } (\bibinfo {year} {2005})}\BibitemShut {NoStop}%
\bibitem [{\citenamefont {Knapp}(2006)}]{Knapp-thesis}%
  \BibitemOpen
  \bibfield  {author} {\bibinfo {author} {\bibfnamefont {M.}~\bibnamefont
  {Knapp}},\ }\emph {\bibinfo {title} {Investigations Into The Structure And
  Properties Of Ordered Perovskites, Layered Perovskites, And Defect
  Pyrochlores}},\ \href@noop {} {Ph.D. thesis},\ \bibinfo  {school} {Ohio State
  University} (\bibinfo {year} {2006})\BibitemShut {NoStop}%
\bibitem [{\citenamefont {Kambe}\ \emph {et~al.}(2002)\citenamefont {Kambe},
  \citenamefont {Ichimaru}, \citenamefont {Sato}, \citenamefont {Yoshida},\
  and\ \citenamefont {Ishii}}]{buckling}%
  \BibitemOpen
  \bibfield  {author} {\bibinfo {author} {\bibfnamefont {S.}~\bibnamefont
  {Kambe}}, \bibinfo {author} {\bibfnamefont {Y.}~\bibnamefont {Ichimaru}},
  \bibinfo {author} {\bibfnamefont {E.}~\bibnamefont {Sato}}, \bibinfo {author}
  {\bibfnamefont {C.}~\bibnamefont {Yoshida}}, \ and\ \bibinfo {author}
  {\bibfnamefont {O.}~\bibnamefont {Ishii}},\ }\href@noop {} {\bibfield
  {journal} {\bibinfo  {journal} {Bull. Yamagata Univ. (Eng.)}\ }\textbf
  {\bibinfo {volume} {27}} (\bibinfo {year} {2002})}\BibitemShut {NoStop}%
\end{thebibliography}%

\end{document}